\documentclass[aps,pra,twocolumn,showpacs,superscriptaddress]{revtex4-1}
\usepackage{xcolor, graphicx, ulem}
\usepackage{amsmath, amssymb}
\usepackage{hyperref}
\usepackage{textcomp}

\begin{document}

\title{Direct calibration of click-counting detectors}

\author{M. Bohmann}
\email{martin.bohmann@uni-rostock.de}
\affiliation{Arbeitsgruppe Theoretische Quantenoptik, Institut f\"ur Physik, Universit\"at Rostock, D-18051 Rostock, Germany}

\author{R. Kruse}
\email{regina.kruse@uni-paderborn.de}
\affiliation{Integrated Quantum Optics, Applied Physics, University of Paderborn, 33098 Paderborn, Germany}

\author{J. Sperling}
\affiliation{Clarendon Laboratory, University of Oxford, Parks Road, Oxford OX1 3PU, United Kingdom}

\author{C. Silberhorn}
\affiliation{Integrated Quantum Optics, Applied Physics, University of Paderborn, 33098 Paderborn, Germany}

\author{W. Vogel}
\affiliation{Arbeitsgruppe Theoretische Quantenoptik, Institut f\"ur Physik, Universit\"at Rostock, D-18051 Rostock, Germany}

\begin{abstract}
	We introduce and experimentally implement a method for the detector calibration of photon-number-resolving time-bin multiplexing layouts based on the measured click statistics of superconducting nanowire detectors. 
	In particular, the quantum efficiencies, the dark count rates, and the positive operator-valued measures of these measurement schemes are directly obtained with high accuracy.
	The method is based on the moments of the click-counting statistics for coherent states with different coherent amplitudes.
	The strength of our analysis is that we can directly conclude---on a quantitative basis---that the detection strategy under study is well described by a linear response function for the light-matter interaction and that it is sensitive to the polarization of the incident light field.
	Moreover, our method is further extended to a two-mode detection scenario.
	Finally, we present possible applications for such well characterized detectors, such as sensing of atmospheric loss channels and phase sensitive measurements.
\end{abstract}

\date{\today}

\maketitle

\section{Introduction}
	For successful implementations of upcoming quantum technologies, e.g., quantum computing~\cite{Nielsen}, quantum metrology~\cite{Giovannetti2011}, or quantum communication~\cite{Gisin2007}, it is crucial to have a deep understanding of the involved quantum processes.
	This requires a profound knowledge of the used measurement devices that are employed to reveal and exploit the quantum features of an experimentally generated state.
	However, no detector is known in quantum optics that allows one to perfectly resolve the photon-statistics of a light field.

	For this reason, in the regime of a few photons, quasi-photon-number-resolving detectors \mbox{(qPNRDs)} have gained major importance.
	One element of these devices is the on/off detector, which has only a binary outcome.
	It either produces a ``click'' if photons are absorbed or, otherwise, remains silent.
	Examples are avalanche photodiodes in the Geiger mode and super-conducting nanowire detectors. 
	Experimentally, it is favorable to employ superconducting nanowire detectors as they achieve a high quantum efficiency~\cite{NTH12}.
	Another key element of a \mbox{qPNRD} is an optical system that equally distributes the impinging photons to several such on/off detectors.
	This can be implemented in various ways~\cite{ABA10, PDFEPW10, WDSBY04, DBJVDBW14, MMDL12, DYSTS11, DBIMHCE13}. 
	In order to efficiently implement a \mbox{qPNRD}, one also uses time-multiplexed detectors~\cite{FJPF03, ASSBW03, RHHPH03} as a resource-saving realization.
	The combination of the optical splitting and the on/off detectors characterizes a click-counting device.
	With the help of a closed-formula description of such \mbox{qPNRDs}~\cite{SVA12}, several quantum effects have been successfully identified in theory and experiment, solely based on the measured click statistics~\cite{SVA12a, SVA13, BDJDBW13, SBVHBAS15, HSPGHNVS15, SBDBJDVW16}. 
	 
	The key quantity which characterizes a detector is the detector response function~\cite{KK64,VW06}.
	It basically describes the transfer from the properties of the incoming radiation field to the detector output signal.
	Especially, it accounts for the light-matter interaction within the device.
	Therefore, it renders it possible to relate the impinging radiation field and the produced detector signal.
	Once the detector response function is determined, it therefore allows for the characterization of arbitrary incident light fields with this detector.
	Despite this importance for quantum optical experiments, the direct measurement of the response function is typically not considered in the existing literature.
	Closing this gap is one of the aims of this paper.

	In order to fully characterize optical measurement devices, different methods have been studied.
	One possible approach is to apply general detector tomography techniques~\cite{tomography1, tomography2, tomography3, tomography4}, e.g., to access properties of \mbox{qPNRDs}~\cite{tomography4, tomography5, tomography6, tomographyqPNRD1, tomographyqPNRD2, tomographyNW1, tomographyNW2, tomographyNW3}.
	Those methods use well-known and controlled input states in order to numerically reconstruct the positive operator-valued measure (POVM) of the detector.
	This approach is universal, as it does not assume physical models of the detector.
	That is, the measurement device is treated as a black box. 
	However, the application of detector tomography bears the intrinsic problem of an inversion from a finite set of measurement outcomes to the POVM, which is acting on an infinite Hilbert space that describes the radiation field.
	This is an ill-defined problem.
	Hence, systematic uncertainties of such an approach have to be propagated along with numerical errors.
	Another way for retrieving the detector response is the use of so-called twin beams~\cite{twinbeam1, twinbeam2,twinbeam3,twinbeam4}.
	They exploit the correlated photon statistics of such states and, therefore, can be seen as a generalization of the method by Klyshko~\cite{Klyshko} for single-photon detectors~\cite{twinbeam2}.
	This technique has the drawbacks that the correlations between the individual beams--one or both being probed by unknown detection systems--have to be well defined and prepared with high accuracy.
	Furthermore, the photon statistics, on which the method is based, cannot be measured; instead, one experimentally obtains the corresponding click-counting statistics.
	However, an inversion from the click-counting statistics to the photon-number statistics suffers from a systematic error which scales with one over the number detection bins~\cite{SVA12}.
	In the case of eight bins, this yields already a systematic error of above 12\%.

	As mentioned above, the need of properly characterized detection devices is vital for the application of quantum light.
	For instance, the reliable generation of quantum states for quantum information tasks requires the knowledge of the detector response function~\cite{SVA14}.
	Furthermore, some free-space communication scenarios demand a transmission of quantum light together with the monitoring of the turbulence of the atmosphere~\cite{Usenko}.
	For example, one can send a quantum signal in one polarization mode and a classical reference in the perpendicular polarization~\cite{Elser,Semenov2012}.
	Thus, it would be also beneficial if one can perform a polarization dependent sensing of a random loss media with the same detection device.
	Additionally, the well-characterized \mbox{qPNRD} systems can be used to perform phase sensitive measurements~\cite{SVA15,LuisSV15,LipferSV15}.
	With such setups, the quantum properties of the radiation field can be directly revealed.

	In this paper, we introduce an efficient technique to characterize \mbox{qPNRDs} and directly determine their detector response function based on moments of the measured click-counting statistics. 
	Our method places only minimal assumptions on the detector and a set of measurements with power-controlled coherent light.
	In particular, no truncation of the Fock space or inversion from the click to the photon statistics is needed.
	From a regression of the obtained data, we can infer the detector characteristics such as quantum efficiencies and dark count probabilities. 
	From this analysis, we can also immediately extract the POVM elements of the detector circumventing the difficulties stemming from ill-posed problems, e.g., a truncation of the Fock space as is needed for detector tomography. 
	We test the method with a two-mode time-bin multiplexing layout and superconducting nanowire on/off detectors.
	Besides the easy applicability of our method, the obtained results yield an accuracy which is as good as the best results reported for the experimentally elaborated twin-beam based method.
	Moreover, we investigate the polarization dependency of the two detector modes and discuss their behavior.
	Finally, we present two applications of such a well characterized \mbox{qPNRD}:
	first, how they can be used for sensing the properties of turbulent atmospheric channels based on a similar theoretical approach and, second, how such detectors systems can be utilized in order to perform phase-sensitive measurements.

	The paper is structured as follows.
	In Sec.~\ref{sec:method} we introduce the calibration method based on the click-counting statistics.
	The used experimental setup is described in Sec.~\ref{sec:experiment}.
	In Sec.~\ref{sec:analysis} we apply our method to the experimental data and investigate the polarization dependence of the detector response function.
	Possible experimental imperfections due to polarization effects are discussed in Sec.~\ref{sec:exp_discussion}.
	Applications of well characterized \mbox{qPNRDs} are given in Sec.~\ref{sec:application}.
	Finally, we summarize and conclude in Sec.~\ref{sec:conclusion}.

\section{Click-moment based detector characterization}\label{sec:method}
	In this section, we describe the theoretical technique to calibrate \mbox{qPNRDs}. 
	Our aim is to infer a detector response function $\hat\Gamma$ that contains the dependency of the detector response (bulk matter of the on/off detector) on the photon number of the incident light field (described by the photon-number operators $\hat{n}$).
	In doing so, we will be also able to retrieve the detection efficiency $\eta$ and the dark count rate $\nu$.
	Therefore, we briefly summarize the theoretical description of \mbox{qPNRDs} and show how we can extract the detector characteristics and the corresponding POVM elements from measured click statistics.
	
	\begin{figure}[h]
		\includegraphics[width=.55\columnwidth]{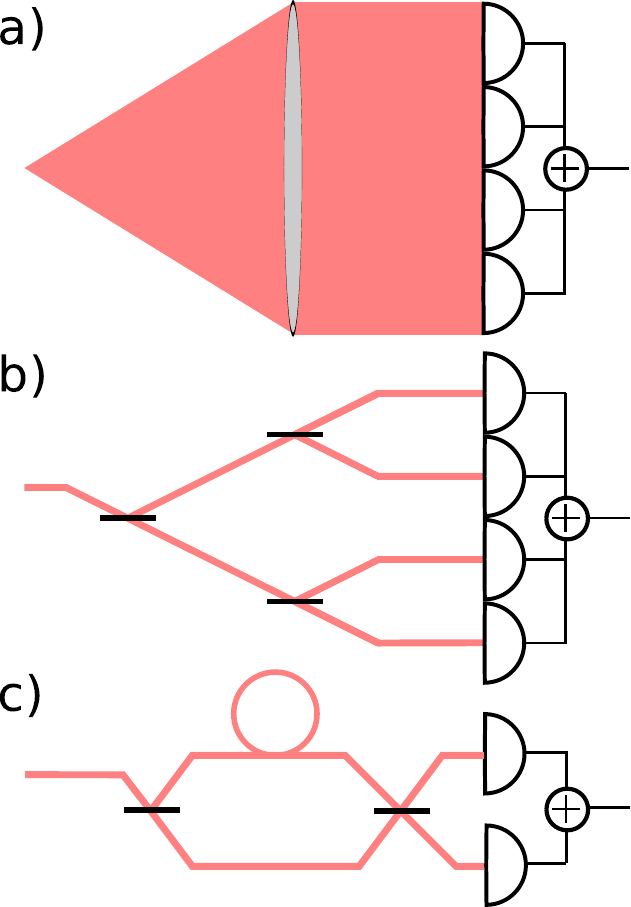}
		\caption{
			(Color online)
			Possible implementations of \mbox{qPNRDs}.
			Three different architectures of \mbox{qPNRDs} with $N=4$ detection bins are shown.
			Each resulting bin is recorded with on/off detectors, the sum of clicks of which yields our desired click-counting statistics.
			(a) In the detector array scenario, an array of on/off detectors is equally illuminated.
			(b) A spatial multiplexing setup is shown in which the incident light is equally divided by multiple 50:50 beam splitters.
			(c) A time-bin multiplexing setup is illustrated, which resembles our implementation.
		}\label{fig:DetectionSchemes}
	\end{figure}

	Before we discuss click-counting detection and introduce our calibration method, let us consider different architectures of \mbox{qPNRD} to which our method applies and which are schematically shown in Fig.~\ref{fig:DetectionSchemes}.
	This includes equally illuminated array detectors [Fig.~\ref{fig:DetectionSchemes}a)], spatial multiplexing [Fig.~\ref{fig:DetectionSchemes}b)], and time-bin multiplexing [Fig.~\ref{fig:DetectionSchemes}c)] (see , e.g., Refs.~\cite{ABA10, DBJVDBW14}, ~\cite{WDSBY04,HSPGHNVS15}, and ~\cite{PDFEPW10, DBIMHCE13} for their according implementations).
	All these realizations have in common that the incident light field is equally split into $N$ different bins ($N=4$ in Fig.~\ref{fig:DetectionSchemes}) and the light in each bin is subsequently recoded with an on/off detector.
	The on/off detectors themselves can be avalanche photodiodes in the Geiger mode or superconducting nanowire detectors.
	The latter ones are employed in our experiment.
	Note that such \mbox{qPNRDs} schemes are frequently used in quantum optical experiments~\cite{ABA10, PDFEPW10, WDSBY04, DBJVDBW14, MMDL12, DYSTS11, DBIMHCE13,FJPF03, ASSBW03, RHHPH03,BDJDBW13, SBVHBAS15, HSPGHNVS15, SBDBJDVW16}.

	Throughout this paper, we mainly deal with two separated click-counting detector systems--labeled as $A$ and $B$--each consisting of $N_{j}$ ($j=A,B$) on/off detectors or time bins.
	However, the treatment can easily be extended to any number of detectors or relaxed to a single one~\cite{SVA13}.
	Then, the system under study is described by a joined click-counting probability $c_{k_A,k_B}$, where we have $k_A$ clicks from the first detector system and simultaneously $k_B$ clicks from the second one ($0\leq k_j\leq N_j$).
	The normalization reads $\sum_{k_A=0}^{N_A}\sum_{k_B=0}^{N_B}c_{k_A,k_B}=1$. 
	The single-mode marginals of the joint click-counting statistics are obtained by $c_{k_A}=\sum_{k_B}c_{k_A,k_B}$ and $c_{k_B}=\sum_{k_A}c_{k_A,k_B}$. 
	For detection systems with equal splitting ratios, the corresponding normalized click-counting statistics follows the quantum version of a binomial distribution~\cite{SVA12,SVA13},
	\begin{align}\label{eq:click_statistics}
	\begin{aligned}
		c_{k_A,k_B}=&\langle{:} \binom{N_A}{k_A}\hat m_A^{N_A-k_A}(\hat 1_A-\hat m_A)^{k_A}\\
		&\times\binom{N_B}{k_B}\hat m_B^{N_B-k_B}(\hat 1_B-\hat m_B)^{k_B}{:}\rangle,
	\end{aligned}
	\end{align}
	where ${:}\,\cdot\,{:}$ indicates the normal-ordering prescription.

	The operators $\hat m_j$, the normally ordered expectation values of which yield the no-click probabilities, are given by
	\begin{equation}
		\hat m_j=e^{-\hat\Gamma_j},
	\end{equation}
	where $\hat\Gamma_j=\Gamma_j(\hat n^{\rm H}_j/N_j,\hat n^{\rm V}_j/N_j)$ is the sought detector response function operator~\cite{SVA13}.
	Note that $\hat\Gamma_j$ depends on the photon-number operator $\hat n_j^{\rm H}$($\hat n_j^{\rm V}$) for the horizontal(vertical) polarization.
	A typical example is a linear response function $\Gamma_j$, i.e.,
	\begin{align}\label{eq:linear}
		\hat \Gamma_j
		=\Gamma_j\left(\frac{\hat n^{\rm H}_j}{N_j},\frac{\hat n^{\rm V}_j}{N_j}\right)=\frac{\eta^{\rm H}_j\hat n^{\rm H}_j}{N_j}+\frac{\eta^{\rm V}_j\hat n^{\rm V}_j}{N_j}+\nu_j,
	\end{align}
	with $\eta^{\rm H/V}_j$ and $\nu_j$ denoting the quantum efficiency and dark count probability per click, respectively.
	Note that we will write $\Gamma_j(\hat n_j/N_j)$ if only one polarization component is considered.
	
	In the following, we describe how we infer the detector response function $\Gamma_j$ directly from the click statistics. 
	For this reason, we consider normally ordered moments of the operators $\hat m_j$
	\begin{align}\label{eq:moments}
		\langle{:}\hat m_A^{l_A}\hat m_B^{l_B}{:}\rangle=\langle{:} e^{-l_A\hat\Gamma_A}e^{-l_B\hat\Gamma_B}{:}\rangle,
	\end{align}
	with $l_j=0,\dots,N_j$ for $j=A,B$. 
	They are obtained from the click statistics via the sampling formula~\cite{SVA13}
	\begin{align}\label{eq:sample}
		\langle{:}\hat m_A^{l_A}\hat m_B^{l_B}{:}\rangle=\sum_{k_A=0}^{N_A-l_A}\sum_{k_B=0}^{N_B-l_B}
		\frac{\binom{N_A-k_A}{l_A}\binom{N_B-k_B}{l_B}}{\binom{N_A}{l_A}\binom{N_B}{l_B}}c_{k_A,k_B}.
	\end{align}
	
	Let us consider any two-mode quantum state in a given polarization which can be written in the Glauber-Sudarshan $P$ function \cite{Sudarshan,Glauber} as
	\begin{align}\label{eq:P-function}
		\hat \rho=\int d^2\alpha \,d^2\beta\, P(\alpha,\beta)|\alpha\rangle\langle\alpha|\otimes |\beta\rangle\langle\beta|.
	\end{align}
	Using this representation, we can now evaluate the expectation value $\langle{:}\hat m_A^{l_A}\hat m_B^{l_B}{:}\rangle$ [see Eq.~\eqref{eq:moments}].
	Due to the normal ordering of the expectation value, the photon-number operators $\hat n_j$ can be replaced by the absolute square of the coherent amplitudes and we obtain 
	\begin{align}\label{eq:P-moments}
	\begin{aligned}
		\langle{:}\hat m_A^{l_A}\hat m_B^{l_B}{:}\rangle=&\int d^2\alpha\, d^2\beta\, P(\alpha,\beta)\\
		&\times e^{-l_A\Gamma_A( |\alpha|^2/N_i)}e^{-l_B\Gamma_B( |\beta|^2/N_i)}.
	\end{aligned}
	\end{align}
	From Eq.~\eqref{eq:P-moments} we see that it is sufficient to know the detector response functions $\Gamma_j$ in dependence on the coherent amplitude in order to determine the expectation value for any quantum state given in the form of Eq.~\eqref{eq:P-function}.
	Therefore, it is sufficient to consider coherent states to determine the response function from the moments.
	For a two-mode coherent state $|\alpha,\beta\rangle$ in a given polarization Eq.~\eqref{eq:P-moments} reduces to
	\begin{align}\label{eq:expectationvalue}
		\langle{:}\hat m_A^{l_A}\hat m_B^{l_B}{:}\rangle= e^{-l_A\Gamma_A( |\alpha|^2/N_i)}e^{-l_B\Gamma_B( |\beta|^2/N_i)}.
	\end{align}
	This is an important relation because it connects the sampled moments with the detector response.
	If we now choose either $l_A=0$ or $l_B=0$, Eq.~\eqref{eq:expectationvalue} reduces to the single-mode form
	\begin{align}\label{eq:singleexp}
		\langle{:}\hat m_j^{l_j}{:}\rangle=e^{-l_j\Gamma_j(|\gamma|^2/N_j)},
	\end{align}
	with $\gamma=\alpha$ and $\beta$ for $j=A$ and $B$, respectively.
	Moreover, selecting the coherent light field in either horizontally or vertically polarization, we can study polarization specific properties of the detector response.
	
	This allows us to relate the detector response function $\Gamma_j$ to the measured moments $\langle{:}\hat m_j^{l_j}{:}\rangle$ and, by rewriting Eq.~\eqref{eq:singleexp}, we directly get an expression for the detector response function:
	\begin{align}\label{eq:Gamma}
		\Gamma_j\left(\frac{|\gamma^{\rm H}|^2}{N_j},\frac{|\gamma^{\rm V}|^2}{N_j}\right)=-\frac{1}{l_j}\ln(\langle{:}\hat m_j^{l_j}{:}\rangle),
	\end{align}
	where the coherent amplitude $\gamma$ is decomposed into its horizontally(vertically) polarized part $\gamma^{H(V)}$.
	It is also worth mentioning that
	$\langle{:}\hat m_A^{l_A}\hat m_B^{l_B}{:}\rangle$ is less than $1$ for all $l_A=l_B\neq0$.
	This follows from the fact that the click-counting statistics for coherent light is a true binomial one~\cite{SVA12}, $c_{k_A,k_B}=\prod_{j=A,B}\left[\binom{N_j}{k_j}p_j^{N_j-k_j}(1-p_j)^{k_j}\right]$, with $0\leq p_j\leq 1$, and the moments can be written as $\langle{:}\hat m_A^{l_A}\hat m_B^{l_B}{:}\rangle=\prod_{j=A,B}p_j^{l_j}$.
	Hence, we immediately observe that $\Gamma_j$, as given in Eq.~\eqref{eq:Gamma}, is always positive.
	
	Applying Eq.~\eqref{eq:Gamma}, we can determine the absolute functional behavior of $\Gamma_j(x)$. 
	By performing a series of measurements with coherent states of different known amplitudes $\{|\gamma_n|\}_n$, we infer the corresponding set of values for the response function $\{\Gamma_j(|\gamma_n|^2/N_j)\}_n$. 
	Via an appropriate regression of this data set, $\{(|\gamma_n|^2,\Gamma_j(|\gamma_n|^2/N_j)\}_n$, we can directly estimate the detector response function. 
	Now, the detector is fully characterized as, in particular, its POVMs are given by
	\begin{align}\label{eq:POVM}
	\begin{aligned}
		\hat\Pi_{k_A,k_B}=\,:&\binom{N_A}{k_A}e^{{-(N_A-k_A})\hat\Gamma_A}(\hat 1_A-e^{-\hat\Gamma_A})^{k_A}\\
		&\times\binom{N_B}{k_B}e^{{-(N_B-k_B)\hat\Gamma_B}}(\hat 1_B-e^{-\hat\Gamma_B})^{k_B}:\, ,
	\end{aligned}
	\end{align}
	see also Eq.~\eqref{eq:click_statistics}.
	Let us emphasize that once the detector response is determined, it gives us the possibility to estimate the POVMs [Eq. \eqref{eq:POVM}], which are applicable to any kind of quantum state.
	While it is possible to utilize $N_j$ ($j=A,B$) different ways to characterize each detection system [$l_j$ values in Eq. \eqref{eq:Gamma} between $1$ and $N_j$], we will only consider the first moment ($l_j=1$) of the click statistics.
	The reason is that the statistical significance of higher-order click moments is typically lower than those of the first moment.

	The main strength of our calibration method is that it is very resource efficient, both from the experimental and the theoretical point of view. 
	We only need to perform a phase-insensitive measurement with power-controlled coherent states.
	We require minimal additional knowledge of the detector system, i.e., that it behaves as a click-counting device \cite{SVA12}.
	On this basis we directly estimate the physical characteristics of the detector system.
	From Eq. \eqref{eq:linear}, we can infer the quantum efficiency and the dark count probability.
	General detector tomography methods do not yield this information, as the detector is treated as a black box.
	Only when giving up the generality of the approach by applying less general detector models \cite{tomography6}, this information can be extracted.

\section{Experimental setup}\label{sec:experiment}
	For the particular detection scenario at hand, we are interested in the polarization dependence of superconducting nanowire detectors.
	This is due to the inner geometry of this on/off detector.
	That is, the wires themselves are aligned mostly in parallel to each other~\cite{NTH12}.
	Hence, the orientation of our two click detectors are set up in such a way that they measure the horizontal and vertical polarization with maximal and minimal efficiency, respectively.

	\begin{figure}[t]
		\includegraphics[width=.9\columnwidth]{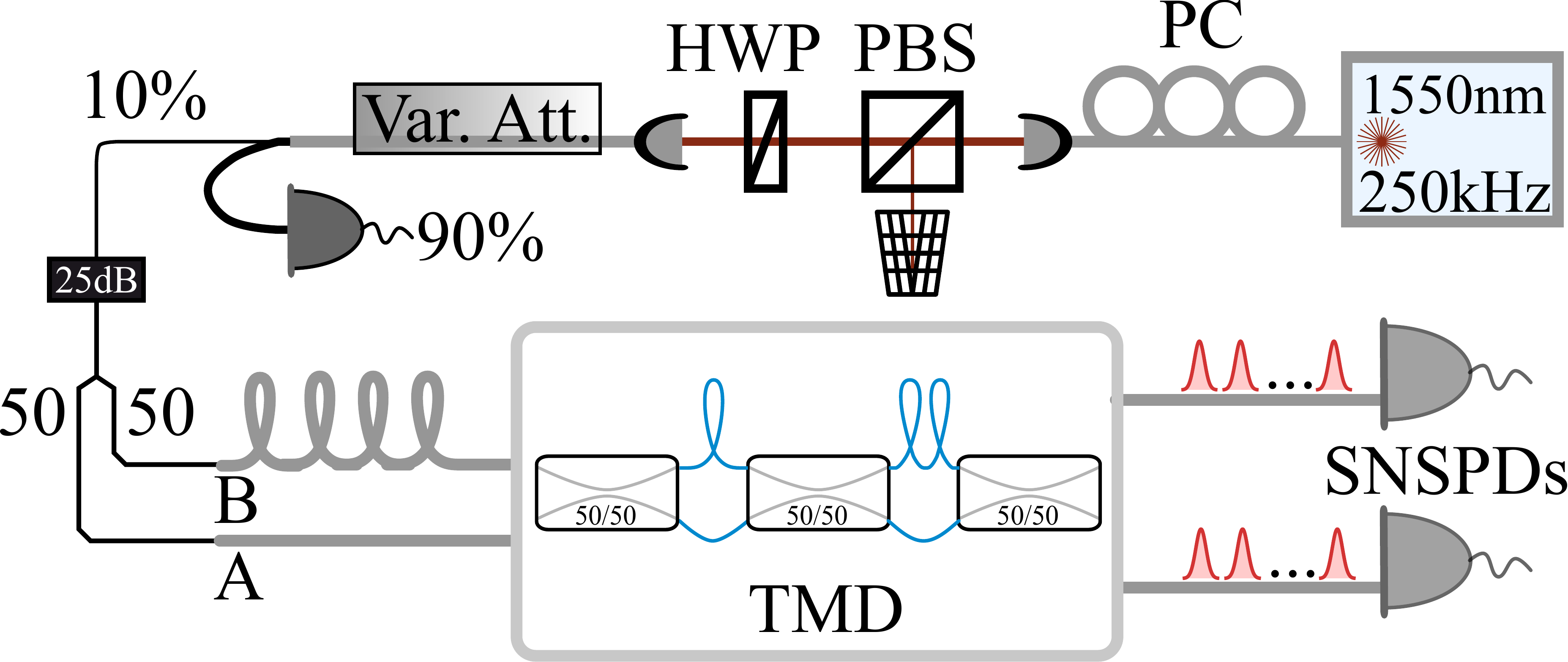}
		\caption{
			Schematic of the experimental setup. 
			Laser pulses at 1550$\,$nm are polarization cleaned and two different polarization states are fed into a fiber-based variable attenuator (Var. Att.).
			The attenuated power is referenced by a 90\% tap-off with a power meter.
			Afterward, the pulses are further attenuated by 25$\,$dB and split by a 50:50 coupler. 
			Finally, they enter the time-multiplexed detection scheme consisting of an eight-bin time-multiplexed fiber-loop detector (TMD) and two superconducting nanowire detectors (SNSPDs).
		}\label{fig:setup}
	\end{figure}

	In Fig.~\ref{fig:setup}, a schematic overview of the experimental setup is given.
	To obtain experimental data for the detector calibration, we send $35$-ps pulses with a repetition rate of 250$\,$kHz to a polarization control (PC) and a polarizing beam splitter (PBS) to clean the polarization of our impinging pulses at a wavelength of 1550$\,$nm.
	After that, we control the polarization via a half-wave plate (HWP) and launch the pulses into a single-mode fiber (SMF-28) network. 
	The action of a variable attenuator (Var. Att.), that decreases the laser power by 0.2$\,$dB every 50$\,$s, is monitored at a 90\% tap-off with a power meter.
	Before the arrival at the time-multiplexed detector (TMD), the pulses undergo further $25$-dB attenuation.
	Then, they are split at a 50:50 coupler. 
	Finally, the pulses impinge on the eight-bin TMD and are detected by superconducting nanowire detectors (SNSPDs). 
	To minimize the influence of the input polarization, the TMD is built from polarization-maintaining single-mode fibers.
	However, as the attenuation components and beam splitter in front of the TMD are \textit{not} polarization maintaining, we still see a polarization mixing effect in the TMD that affects the detection efficiencies of the different time bins.
	We discuss the impact of experimental imperfections on the results and the method in Sec.~\ref{sec:exp_discussion}.

\section{Application of the method}\label{sec:analysis}
	\begin{figure*}
		\includegraphics[width=0.85\columnwidth]{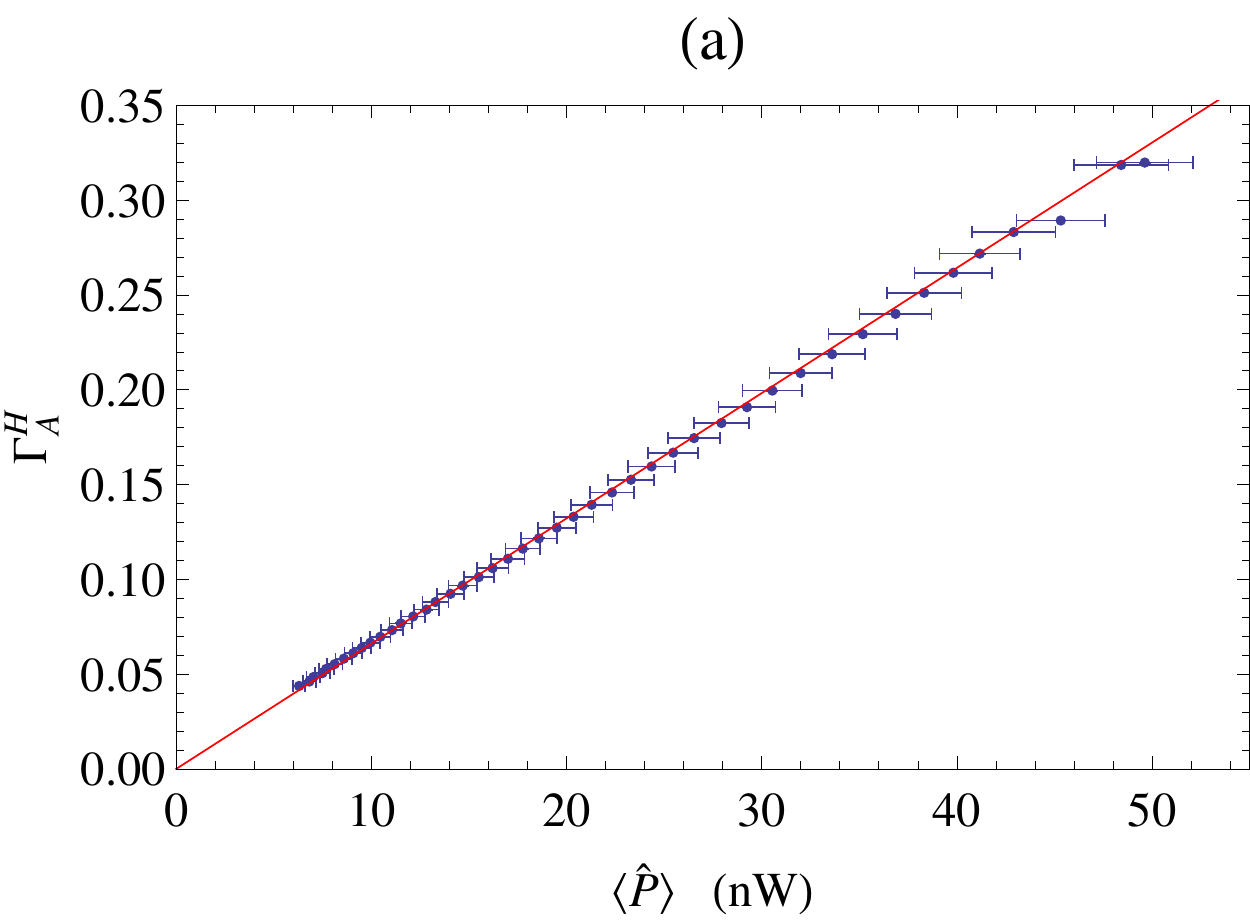}
		\hspace{1cm}
		\includegraphics[width=0.85\columnwidth]{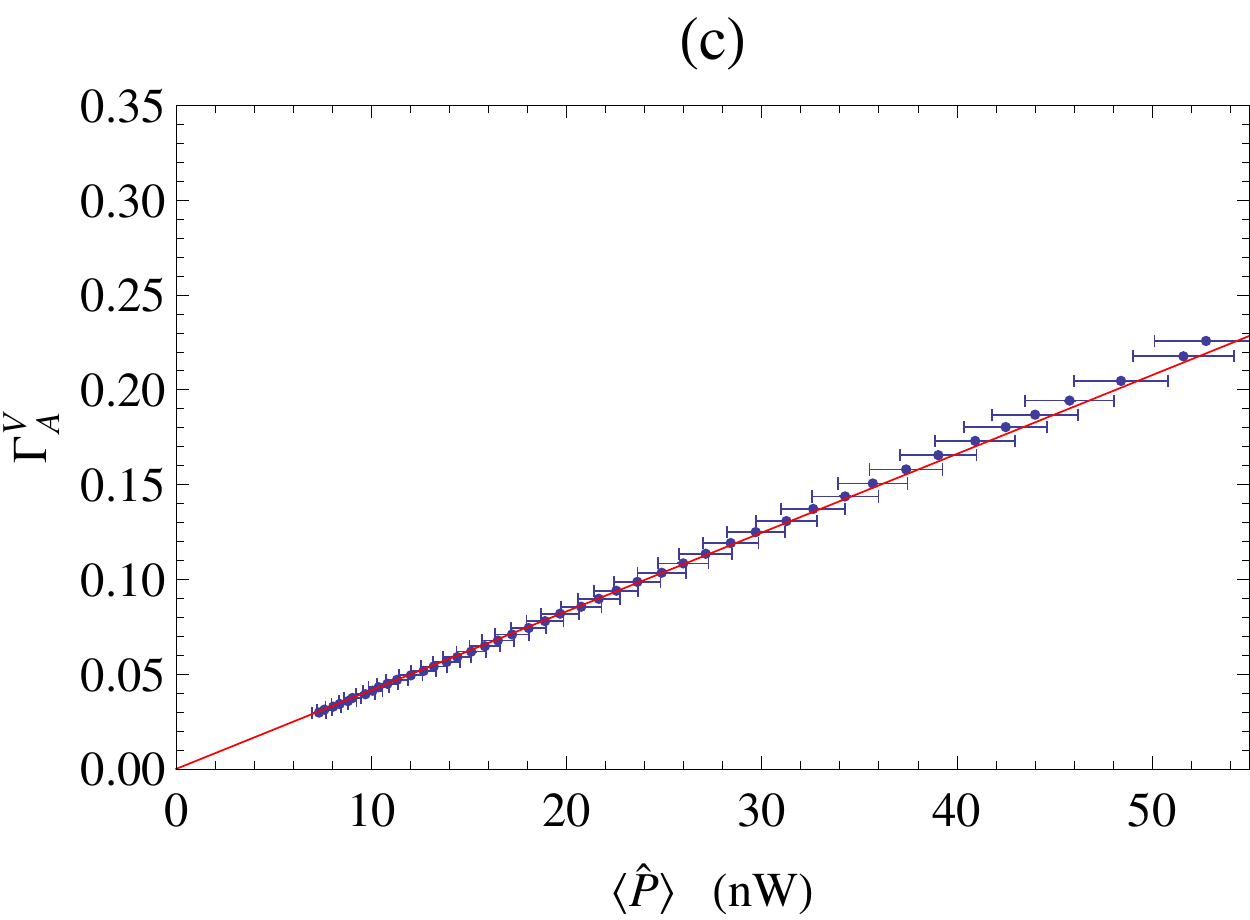}
		\\\vspace{0.1cm}
		\includegraphics[width=.85\columnwidth]{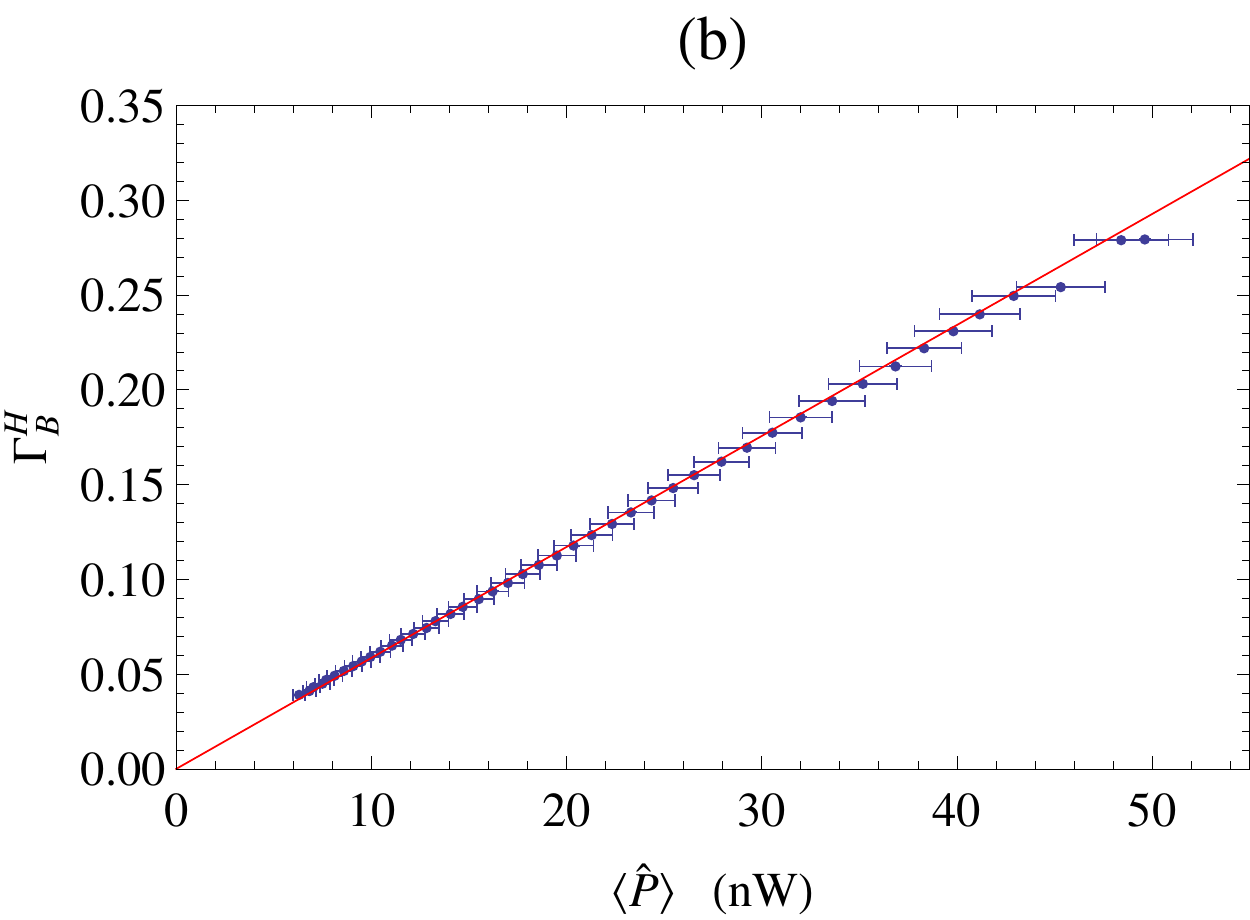}
		\hspace{1cm}
		\includegraphics[width=.85\columnwidth]{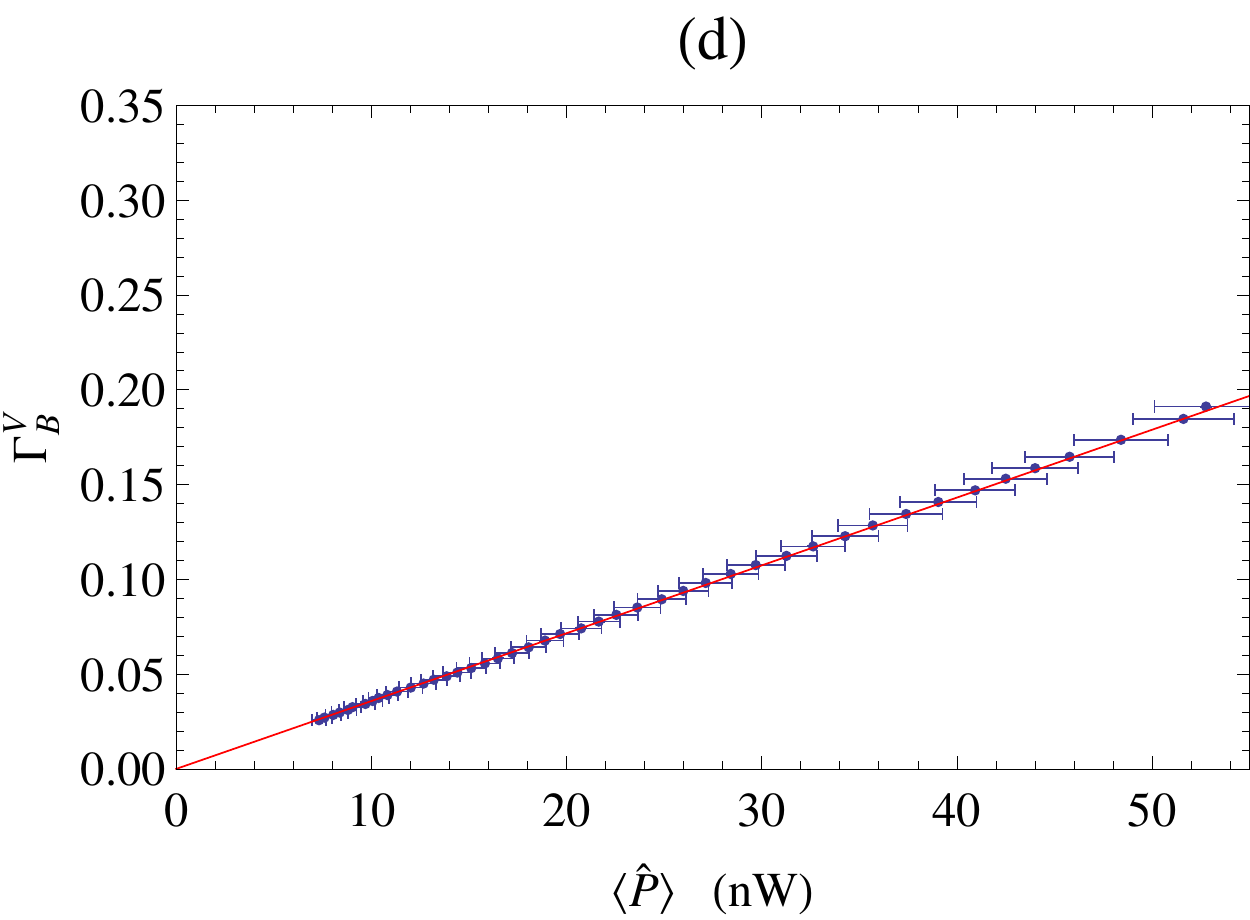}
		\caption{
			Detector responses for horizontal and vertical polarization. 
			(a) Measured data points including error bars for the detector response over the measured intensities (blue points) and the linear regression for $\Gamma^H_A$ (red line). 
			(b) Analysis of detector $B$ depicting the regression of  $\Gamma^H_B$.
			(c) Regression of $\Gamma^V_A$.
			(d) Regression of $\Gamma^V_B$.
		}\label{fig:pAB}
	\end{figure*}

	In this section, we perform a full characterization of the two-mode TMD system described above with the use of the method introduced in Sec.~\ref{sec:method}.
	We first determine the detector response for both, horizontally and vertically polarized light, as the nanowire detectors show a polarization dependent quantum efficiency~\cite{tomographyNW2}.
	Furthermore, we extract the detector characteristics such as the quantum efficiency, as well as the POVMs of the detector.
	Additionally the response functions for intermediate polarizations are determined.

\subsection{Reconstruction of the response function}
	First, we show how to obtain the click-counting statistics and its moments from the measured data. 
	The experiment delivers the measured event distribution $C_{k_A,k_B}$. 
	Normalizing this distribution by the overall number of events, $C=\sum_{k_A=0}^{N_A}\sum_{k_B=0}^{N_B} c_{k_A,k_B}$, yields the joint click-counting statistics $c_{k_A,k_B}$.
	We directly sample the corresponding moments $\overline{\langle{:}\hat m_A^{l_A}\hat m_B^{l_B}{:}\rangle}$ from the measurement data [cf. Eq.~\eqref{eq:sample}].
	The over-lines indicate the sampled mean values.
	The statistical error (i.e., the standard error of the mean) of the moments is determined by the sample standard deviation, which is given by \cite{SBVHBAS15}
	\begin{equation}
	\begin{aligned}
		&\sigma\left(\overline{\langle{:}\hat m_A^{l_A}\hat m_B^{l_B}{:}\rangle}\right)=\frac{\sigma(\langle{:}\hat m_A^{l_A}\hat m_B^{l_B}{:}\rangle)}{\sqrt{C-1}}=\frac{1}{\sqrt{C-1}}\\
		&\times\sqrt{\sum_{k_A{=}0}^{N{-}l_A} \hspace{-0.1cm} \sum_{k_B{=}0}^{N{-}l_B} \hspace{-0.2cm} c_{k_A,k_B}\hspace{-0.1cm} \left(\hspace{-0.1cm}\frac{\binom{N{-}k_A}{l_A}\binom{N{-}k_B}{l_B}}{\binom{N}{l_A}\binom{N}{l_B}}{-}\overline{\langle{:}\hat m_A^{l_A}\hat m_B^{l_B}{:}\rangle}\hspace{-0.1cm}\right)^2}.
	\end{aligned}
	\end{equation}
	Eventually, we get the estimated moments, $\langle{:}\hat m_A^{l_A}\hat m_B^{l_B}{:}\rangle=\overline{\langle{:}\hat m_A^{l_A}\hat m_B^{l_B}{:}\rangle}\pm\sigma\left(\overline{\langle{:}\hat m_A^{l_A}\hat m_B^{l_B}{:}\rangle}\right)$.
	As we directly extract the moments from the measured click statistics, no additional data processing is needed.
	This allows us to obtain the first required quantity for the characterization based on Eq.~\eqref{eq:Gamma}.

	The second quantity, the intensities $|\gamma_n|^2$, are obtained from the reference power measurement.
	The power meter monitors the power of the coherent laser light which is used for the calibration (see~Fig.~\ref{fig:setup}). 
	For the measured power, we introduce the power operator $\hat P$.
	The corresponding photon-number operator per pulse is
	\begin{align}\label{eq:Pn}
		\hat n_j=\chi\hat P
	\end{align}
	for both modes $j=A,B$, where $\chi=(1.77\pm0.17)\cdot10^8\,\mathrm{W}^{-1}$ is the attenuation factor between the reference tap-off and the power that enters the TMD. 
	The corresponding error in $\chi$ originates from uncertainties of the power meter via error propagation.
	The reference power is recorded with a measurement uncertainty of $\pm5\%$.
	The measured intensities then are given by $\langle \hat n_i\rangle=|\gamma_k|^2$.
	For both polarizations, we record the click-counting statistics for 45 different intensities (powers).

	To retrieve the unknown detector responses $\Gamma_i$, we use Eq.~\eqref{eq:Gamma} with the first moments as argued in Sec.~\ref{sec:method}.
	We plot the obtained values of the response functions $\Gamma_i$ for both modes $A$ and $B$ for horizontally or vertically polarized light in Fig.~\ref{fig:pAB}.
	We depict the data points for the $\Gamma_j$ (blue) over the different incident powers with their measurement uncertainties. 
	We indicate the different polarizations by a superscript of $H$ and $V$ for horizontal and vertical.
	Note that the uncertainties of the obtained $\Gamma_j=-\ln\langle:\hat m:\rangle$, which are determined by the statistical uncertainties of the measured $\Gamma_j$, are so small that they are almost not resolved in the plots (see Fig.~\ref{fig:pAB}).
	Hence, it becomes clear that the dominating errors are the ones originating from the power measurement.

	To infer the functional behavior of the response functions, an appropriate regression of the data points is needed. 
	In order to get a direct relation to the measured quantity, the power $\langle\hat P\rangle$, we will further express the response function in terms of $\hat P$ instead of $\hat n$ using the relation Eq.~\eqref{eq:Pn}.
	This allows us to directly work with the observed physical quantities.
	For applying the regression we expand the response function in a Taylor series,
	\begin{align}
		\Gamma_j(\hat P_j/N_j)=\sum_{t=0}^{\infty}\tilde \Gamma_j^{(t)}(\hat P_j/N_j)^t.
	\end{align}
	The coefficients $\tilde\Gamma^{(0)}$ and $\tilde\Gamma^{(1)}$--neglecting the lower index--are the constant and linear contribution to this expansion, respectively.
	Note that the coefficients which correspond to the expansion in the power $\hat P$ are indicated with a tilde.
	To determine the influence of non-linear contributions, we first use the Taylor expansion up to the third order. 
	This yields, for both polarizations and both modes, a ratio between the quadratic and the linear coefficient $\tilde \Gamma^{(2)}/\tilde \Gamma^{(1)}$ and cubic and linear coefficient $\tilde \Gamma^{(3)}/\tilde \Gamma^{(1)}$ of the order of $10^{-3}$ and $10^{-4}$, respectively.
	Hence, higher order terms can be neglected as the response function is properly described by a linear function only.

	The expansion coefficients $\tilde \Gamma^{(0)}$ and $\tilde \Gamma^{(1)}$ are also related to the dark count rates $\nu$ and the quantum efficiencies $\eta$, respectively; see also Eqs.~\eqref{eq:linear} and~\eqref{eq:Pn}.
	Namely, $\tilde \Gamma^{(0)}$ itself is the dark count rate, $\Gamma^{(0)}=\nu=\tilde\nu$, whereas $\tilde \Gamma^{(1)}$ is the scaled quantum efficiency per nanowatt, $\tilde\Gamma^{(1)}=\chi\eta=\tilde\eta$. 
	Thus, we can determine the coefficients of the linear response
	\begin{align}\label{eq:linResP}
		\Gamma_j(\hat P_j/N_j)=\tilde \eta_j\hat P_j/N_j+\tilde\nu_j.
	\end{align}
	Similar to the higher-order terms we find that \mbox{$\tilde\Gamma^{(0)}_j/\tilde\Gamma^{(1)}_j\approx 0$} and we can perform a regression of the form $f(x)=ax$.
	This means the detectors are dark count free, which agrees with the known behavior of SNSPDs of being virtually dark count free~\cite{SSTT13}.
	We use a weighted total least-squares regression algorithm~\cite{regression}, which takes into account the measurement and statistical uncertainties.
	In Fig.~\ref{fig:pAB}, the resulting linear functions (red lines) are plotted.
	It can be seen that they fit the data properly.
	In addition, Table~\ref{tab:table1} summarizes the results from the linear regressions for the two TMD detectors and the two polarizations.

	\begin{table}[htbp]
		\centering
		\caption{
			Parameters and error estimates ($\sigma$) of the linear detector response in Eq.~\eqref{eq:linResP}.
			}\label{tab:table1}
		\begin{tabular}{cccccc}
		\hline
		$\Gamma$ & $\tilde \eta\,$ $(1/{\rm nW})$ & $\sigma_{\tilde\eta}\,$ $(1/{\rm nW})$ &  $\eta\,[\%]$ & $\sigma_\eta\,[\%]$  & Figure \\
		\hline
		$\Gamma_A^{\rm H}$ & $52.86{\times} 10^{-3}$ & $0.04{\times} 10^{-3}$ &  29.8 & 2.8 & \ref{fig:pAB}(a)\\
		$\Gamma_B^{\rm H}$ & $46.83{\times} 10^{-3}$ & $0.02{\times} 10^{-3}$ &  26.4 & 2.5 & \ref{fig:pAB}(b)\\
		$\Gamma_A^{\rm V}$ & $33.23{\times} 10^{-3}$ & $0.04{\times} 10^{-3}$ &  18.7 & 1.8 & \ref{fig:pAB}(c)\\
		$\Gamma_B^{\rm V}$ & $28.63{\times} 10^{-3}$ & $0.01{\times} 10^{-3}$ &  16.1 & 1.6 & \ref{fig:pAB}(d)\\
		\hline
		\end{tabular}
	\end{table}

	As already discussed above, the detector is fully characterized by its response function; see~\cite{KK64} for photoelectric detection models.
	Using the parameters from Table~\ref{tab:table1}, we evaluate the accuracy of our detector characterization.
	We see that with our method, the slope $\tilde\eta$ can be determined with a relative uncertainty up to $0.04\%$.
	Other approaches to calibrate \mbox{qPNRDs} use twin beams and reach relative uncertainties of $0.18\%$ in Ref.~\cite{twinbeam1}, $0.04\%$~\cite{twinbeam2}, $0.39\%$~\cite{twinbeam3}, and $5\%$~\cite{twinbeam4}. 
	For example, the general detector tomography in~\cite{tomography6} yields a relative error estimated of about $8\%$. 
	Compared to these benchmarks, our approach provides a comparable or even better accuracy requiring only measurements with laser light.
	
	So far, we have considered the detector response in terms of the measured power which let us directly infer the behavior of the detector system in terms of the experimental quantities. 
	Yet, the interpretation of the parameters of the linear response in this representation has another intuitive physical interpretation.
	One might also directly identify the quantum efficiency $\eta$, [see Eq. \eqref{eq:linear}].
	Let us also stress that the quantum efficiencies represent the overall efficiencies of the whole TMD detector system and even account for all losses behind the 50:50 beam splitter in Fig.~\ref{fig:setup} as well as the detection efficiency of the nanowire detectors.
	With this information, we have completely characterized the two-mode TMD system and we can determined the POVMs of the system via Eq.~\eqref{eq:POVM}.
	Using the extracted data from Table~\ref{tab:table1}, we evaluate the accuracy of our detector characterization. 
	From Table~\ref{tab:table1}, we see that with our method the quantum efficiency can be determined with a relative uncertainty of $9.4\%$. 
	The absolute error for the quantum efficiencies is determined and limited by the accuracy of the scaling factor $\chi=(1.77\pm0.17)\cdot10^8\,\mathrm{W}^{-1}$. 
	In our case, the uncertainty in $\chi$ was dominated by the accuracy of the power meter at small powers that arrived at the TMD. 
	The optimization of this accuracy is, however, an experimental issue of the available equipment and does not limit our characterization method in general.

\subsection{Polarization dependency}
	\begin{figure}[t]
		\includegraphics[width=0.85\columnwidth]{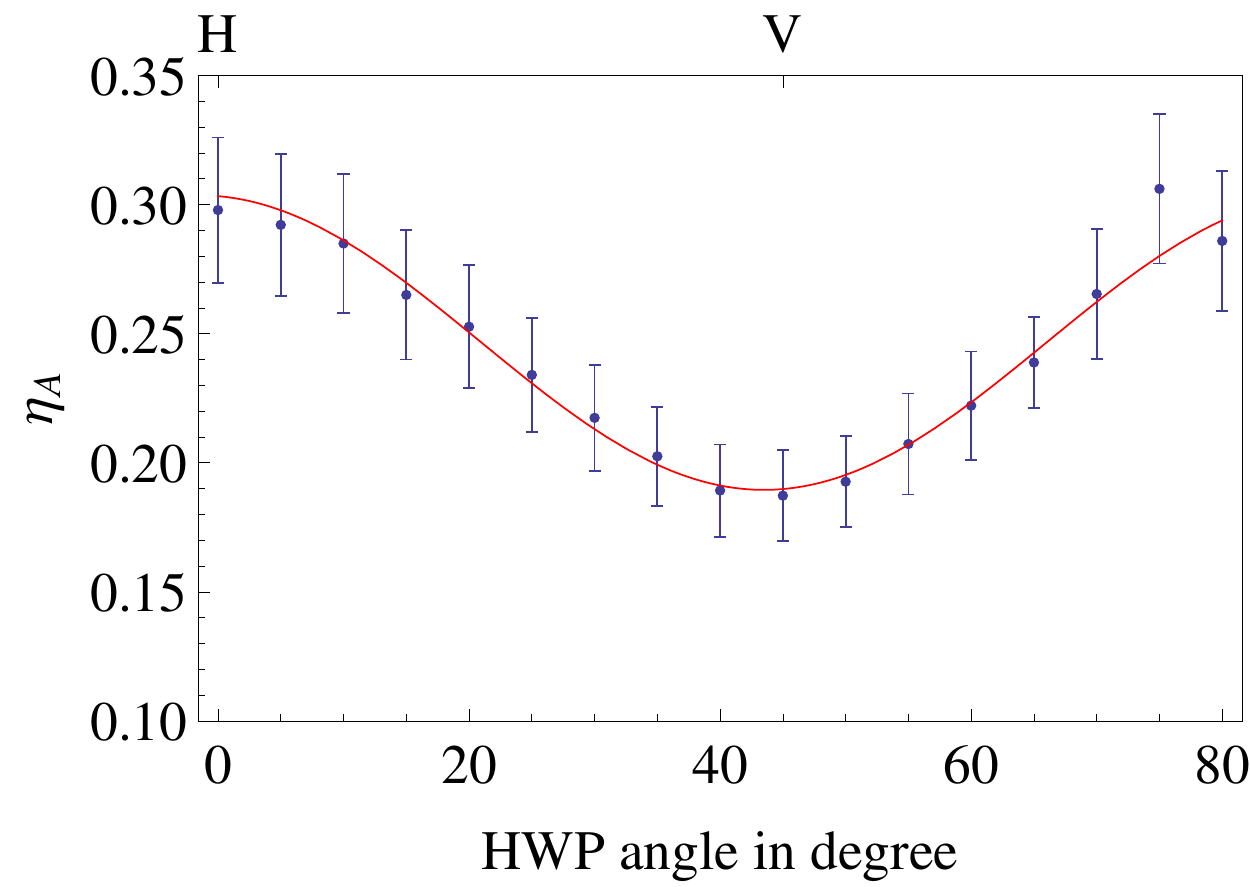}\\\vspace{0.3cm}
		\includegraphics[width=0.85\columnwidth]{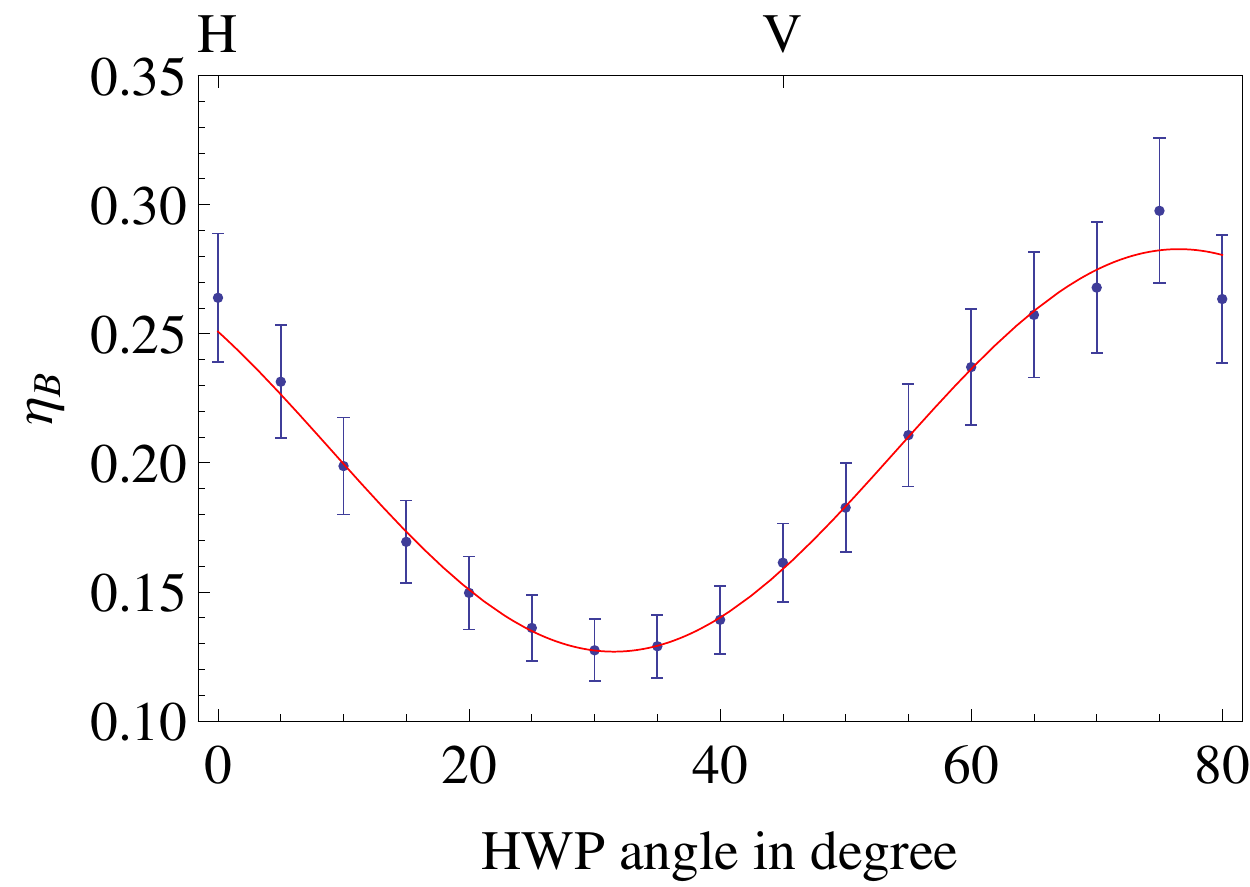}
		\caption{
			The polarization dependency of the determined quantum efficiencies $\eta_A$ (top) and $\eta_B$ (bottom) and their errors.
			The solid lines provide a cosine fit function [cf. Eq.~\eqref{eq:fit}].
			The horizontal and vertical polarizations are indicated.
		}\label{fig:pol}
	\end{figure}

	Let us now study the polarization dependency of the quantum efficiencies for both modes.
	For 17 different polarizations, we recorded the data sets for extracting the detector response in the same way as described above.
	As all response functions show a linear behavior and the detector is virtually dark count free, the comparison of the different polarizations reduce to the comparison of the quantum efficiencies.
	The quantum efficiencies for these mixed polarizations [see the general expression in Eq.~\eqref{eq:linear}] and the errors for both modes are shown in Fig.~\ref{fig:pol}.
	We fit the data with the function
	\begin{align}\label{eq:fit}
		\eta(\phi)=\frac{\eta_{\rm max}-\eta_{\rm min}}{2}\cos[4(\phi+\phi_0)]+\frac{\eta_{\rm max}+\eta_{\rm min}}{2},
	\end{align}
	where $\phi$ is the angle of the HWP in degree and $\eta_{\rm max}$ and $\eta_{\rm min}$ correspond to the maximal and minimal quantum efficiency, respectively.

	For mode $A$ we get $\eta_{A,{\rm max}}=30.2$, $\eta_{A,{\rm min}}=19.3$, and $\phi_{A,0}=1.5^\circ$.
	Especially, $\eta_A$ shows a cosine dependency where positions of the minimum and maximum agree with the horizontal ($H$) and vertical ($V$) polarization as $\phi_{A,0}$ is almost zero.
	This represents the polarization dependency one would expect for such nanowire detectors due to their geometry~\cite{polarisation}.

	In the case of mode $B$ we examine a different behavior and get $\eta_{B,{\rm max}}=28.2$, $\eta_{B,{\rm min}}=12.8$, and $\phi_{B,0}=13.5^\circ$.
	We immediately see that the cosine function is significantly shifted by $\phi_{B,0}$ and, hence, $\eta_{B,{\rm max}}$ and $\eta_{B,{\rm min}}$ do not coincident with the $H$ and $V$ polarization, respectively.
	Additionally, $\eta_{B,{\rm min}}$ is by a factor of $0.66$ smaller than $\eta_{A,{\rm min}}$ while the ration between $\eta_{B,{\rm max}}$ and $\eta_{A,{\rm max}}$ is $0.93$.
	These effects lead to the question of why the two detector modes $A$ and $B$ exhibit such a different behavior.
	Or rather, why mode $A$ shows a comprehensible polarization dependency and why mode $B$ does not.

	To understand this we have to consider the difference in the detection of the detector modes.
	From Fig.~\ref{fig:setup} we see that the only distinction between the detector modes is that the light of mode $B$ first passes an optical fiber, which serves as a delay line, before it enters the detector.
	The overall polarization shift may be explained by a polarization rotation in the delay line.
	However, this is superimposed by polarization mixing effects in the whole TMD as the fibers are only polarization maintaining for the $H$ and $V$ polarization.
	The strength of this effect depends on the length of the fiber the radiation field passes through and, hence, is different for every time bin.
	In the following section we will discuss the polarization effects and how to interpret the obtained results in more detail.
	
\section{Experimental imperfections and impact on the model}\label{sec:exp_discussion}

	In the previous sections, we have introduced and applied the detector calibration by click moments to experimental data. 
	While we have shown that the calibration method is reliable, numerically stable, and yields accurate results, our model is based on assumptions that are not necessarily fulfilled in an experiment. 
	More specifically, we assumed that the click statistics follow a binomial distribution, compare Eq.~\eqref{eq:click_statistics}, which holds only true if the radiation field is equally distributed on the different time-bins and each bin exhibits the same quantum efficiency.
	In an experimental implementation this will not always be the case. 
	In particular, we observed a strong polarization dependency of the detector which one has to account for.
	Therefore, let us consider the distribution of the output signals of the different time bins of the qPNRD.

	\begin{figure}[b]
		\includegraphics[width=.8\columnwidth]{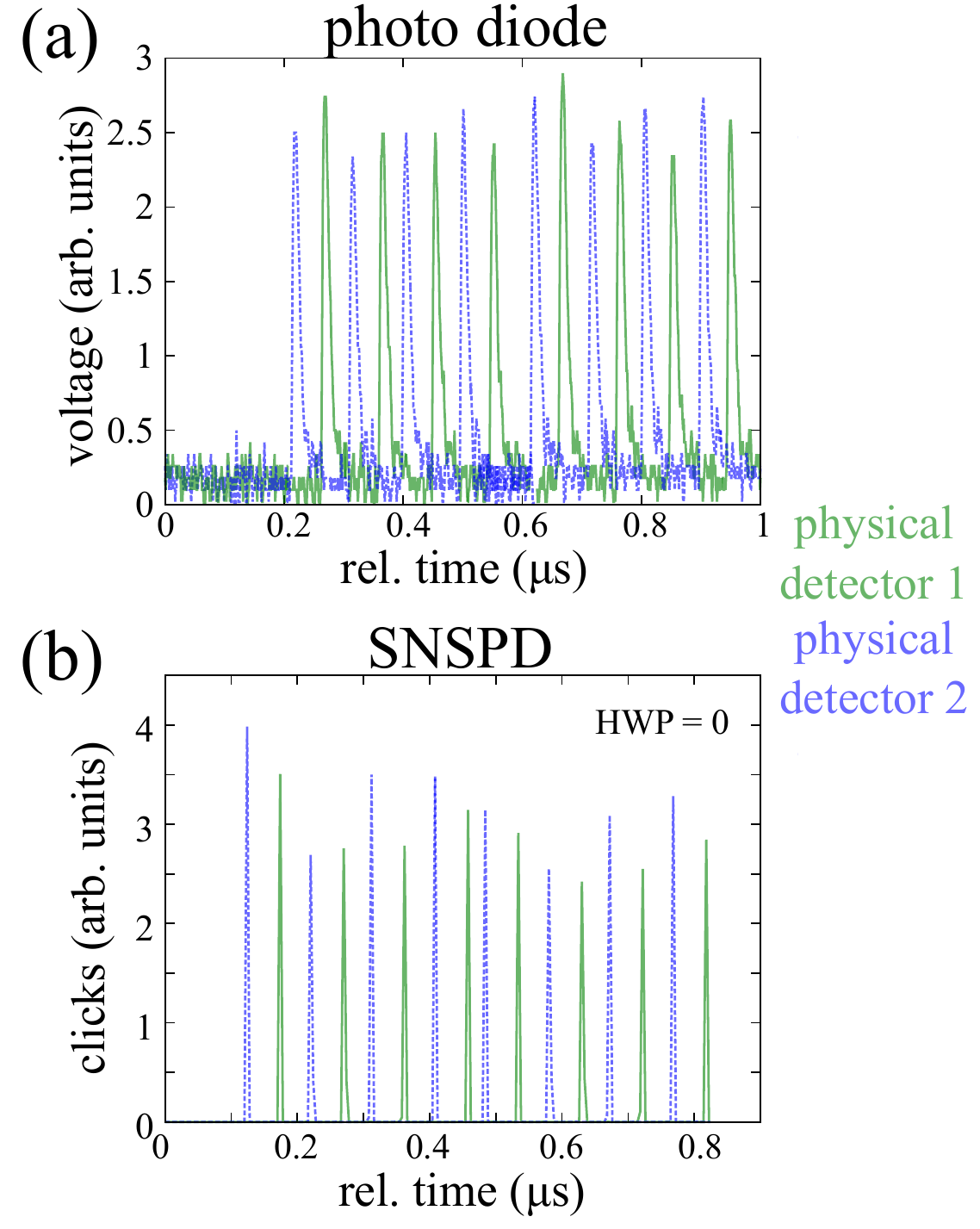}
		\caption{
			Effect of polarization mixing in the TMD. 
			The solid and dashed curves correspond to the two physical detectors.
			In (a), we have measured the direct splitting ratios of the TMD on a polarization insensitive photo diode and plotted the voltage response. 
			While the splitting ratios are not completely even, they are within the specified uncertainty of the used components. 
			In (b) we have plotted the SNSPD time response for a measurement with horizontal polarization. 
			However, the observable imbalance cannot be completely explained by unequal splittings. 
			A time-bin dependent mixing of the polarization state arriving at the detectors leads to uneven detection efficiencies, lowering the overall performance of the TMD.
		}\label{fig:comp_photo_snspd}
	\end{figure}

	In Fig. \ref{fig:comp_photo_snspd}(a), we have depicted the output signal of the time bins by measuring the TMD pulse train on a polarization-insensitive photodiode.
	While some imbalance is observable in the peak heights, it is fairly small and within the fabrication uncertainties of the used 50:50 fiber couplers. 
	Although even the slight imbalance in the peak height distribution will enter a quantifiable error to our method, it will still be relatively small and may be neglected. 
	This situation changes when measuring the splitting ratio with the superconducting nanowire detectors and horizontal polarization in Fig. \ref{fig:comp_photo_snspd}(b). 
	It can be directly seen that the signal ratios of the TMD are drastically changed compared to the polarization insensitive measurement in Fig. \ref{fig:comp_photo_snspd}(a). 
	In contrast to the photodiode measurement in Fig. \ref{fig:comp_photo_snspd}(a), the peak heights in Fig. \ref{fig:comp_photo_snspd}(b) are governed not only by the splitting ratios of the fiber components but also by a polarization mixing effect due to the different fiber lengths of the TMD.
	Then, the light is for each time bin in a different polarization state and this yields different detection efficiencies for each bin.
	Hence, one needs to account for this polarization dependency and possible polarization mixing.

	Concerning the detector calibration, the imbalance of the splitting ratios and possible polarization effects imply that the determined detector response cannot be interpreted as the response for each bin separately anymore, but as the average of all bins of the detector.
	This effects especially the mode $B$ (see Fig.~\ref{fig:pol} and the related discussion).
	Therefore, let us consider an incoming coherent state $|\alpha\rangle$ which is (unequally) split on to $N$ bins $|t_1\alpha,\dots,t_N\alpha\rangle$, with $\sum_i|t_i|^2\leq 1$.
	Note that in a balanced and lossless splitting case $|t_i|^2=1/N$ holds.
	For the general case, we consider $\ln\langle{:}e^{\Gamma_1(\hat n_1)}\dots e^{\Gamma_N(\hat n_N)}{:}\rangle=\sum_{i=1}^N\Gamma_i(|\alpha|^2/|t_i|^2)$.
	By Taylor expanding this quantity, we get the $\overline{\Gamma}(|\alpha|^2)=\sum_{j=0}^\infty \sum_i c_i^{(j)}(|\alpha|^2)^j=\sum_{j=0}^\infty \overline{c^{(j)}}(|\alpha|^2)^j$.
	Here, the overlines denote the average over all bins.
	In particular, one receives the averaged quantum efficiency as $\overline{\eta}=\overline{c^{(1)}}$.
	In this sense, one can characterize a qPNRD in mean, with the restriction that the response of the individual bins cannot be determined.
	Taking the imbalanced splitting and the polarization dependency of the quantum efficiencies by the above treatment in to account, we see that our results in Sec.~\ref{sec:analysis} have to be interpreted in this fashion, i.e., they represent averaged efficiencies of the individual bins.
	Let us stress again that this is no problem in principle.
	It just means that the entire qPNRD setup is characterized, but not its individual components.

	From the experimental point of view, there are solutions to circumvent the inequality of the bins. 
	For polarization independent click detectors, such as avalanche photo diodes, one only has to assure an equal splitting ratio.
	In the polarization dependent case one can rely only on polarization maintaining fiber-integrated components for both attenuators and splitters, or one has to control the polarization directly in front of the TMD.
	This solution is practical, but requires some care to align the polarization correctly in the fiber-integrated network, as the physical splitting ratios of the 50:50 couplers are superimposed with the different detection efficiency.

\section{Application of calibrated click detectors}\label{sec:application}
	We will discuss applications of well characterized \mbox{qPNRDs}.
	By knowing the full detector response, one can use the detectors in order to characterize the losses in an unknown quantum channel.
	Here, we will demonstrate how one can sense the moments of fluctuating loss in atmospheric channels.
	Furthermore, a system of well characterized \mbox{qPNRDs} can be used to realize phase sensitive measurements.
	
\subsection{Sensing the turbulent atmosphere}
	Assume a well-characterized \mbox{qPNRD} with a dark count-free linear response function.
	For the following considerations it is useful to rename $\Gamma(\hat n/N)=\eta_{\rm det}\hat n/N$, with the previously determined efficiency $\eta_{\rm det}$.
	We consider a coherent probe state $|\alpha\rangle$ which first propagates through a turbulent loss channel before it arrives at the qPNRD.

	The action of the turbulent loss is described by an appropriate probability distribution of the transmittance $\mathcal{P}(\eta)$~\cite{Semenov2009,beamwandering,VSV2016}, where $\eta$ is the intensity transmittance of the channel.
	The detected moment for a coherent state then reads as
	\begin{align}\label{eq:tur}
		\langle:e^{-(\eta_{\rm det}\hat n/N)}:\rangle_{\rm tur}=\int_0^1 d\eta\mathcal{P}(\eta)e^{-(\eta\eta_{\rm det}|\alpha|^2/N)}.
	\end{align}
	By a Taylor expansion of $e^{-(\eta\eta_{\rm det}|\alpha|^2/N)}$ in $|\alpha|^2$ around $|\alpha|^2=0$ one can then retrieve the moments of the probability distribution of the transmittance, $\mathcal{P}(\eta)$.
	Note that this Taylor expansion corresponds to an extrapolation of $e^{-(\eta\eta_{\rm det}|\alpha|^2/N)}$ to $|\alpha|^2=0$ and, hence, can be obtained from a set of measurements similar to that described in Sec.~\ref{sec:method}.
	For an expansion of Eq.~\eqref{eq:tur} up to the second order, we get
	\begin{align}
	\begin{aligned}
		&\langle:e^{-(\eta_{\rm det}\hat n/N)}:\rangle_{\rm tur}
		\\\approx&
		1-\eta_{\rm det}\langle \eta\rangle_{\rm tur}|\alpha|^2+\frac{1}{2}\eta_{\rm det}^2\langle \eta^2\rangle_{\rm tur}|\alpha|^4,
	\end{aligned}
	\end{align}
	where $\langle \eta^j\rangle_{\rm tur}=\int_0^1 d\eta\mathcal{P}(\eta)\eta^j$ ($j\in\mathbb{N}$) are the corresponding moments of the transmittance which characterize $\mathcal{P}(\eta)$.
	The properties of atmospheric channels, i.e., the moments of $\mathcal{P}(\eta)$, can be sensed in this way, which is important to identify which nonclassical effects of the radiation field can survive in such channels~\cite{BSSV16,BSSV16b}.
	Such an analysis is the basis for the development of optimal schemes for global quantum communication using atmospheric free-space links.
	
\subsection{Phase sensitive measurements}
	Another application of well-characterized \mbox{qPNRDs} are phase sensitive measurement setups.
	In particular, it is possible to transfer the concepts of balanced and unbalanced homodyne detection to the few-photon regime by using \mbox{qPNRDs} (see~\cite{SVA15} and~\cite{LuisSV15}, respectively).
	Even multiport homodyne detection with \mbox{qPNRDs} has been studied~\cite{LipferSV15}.
	By doing so, phase sensitive features of the quantum state understudy can be examined.
	
	In the case of balanced homodyne click detection, the quantum state understudy is mixed with a phase and amplitude controlled coherent reference state (local oscillator) at a 50:50 beam splitter.
	Subsequently, both output modes are measured with \mbox{qPNRDs}.
	It has been shown that such a setup yields the measurement of a nonlinear quadrature operator~\cite{SVA15}
	\begin{align}
		\hat X(\varphi)=N(\hat m_A-\hat m_B),
	\end{align}
	with $\hat m_j={\rm e}^{-\hat \Gamma_j}$.
	Here $A$ and $B$ denote the two detection modes and $N$ is the number of click detectors.
	We directly see that the generalized quadrature operator $\hat X(\varphi)$ depends on the detector response function.
	Hence, the characterization of the used \mbox{qPNRDs}, i.e. the determination of $\Gamma_{j}$, is crucial for such a measurement setup.
	Specifically, such a characterization, is helpful for the design of a phase sensitive experiment in order to specify conditions under which certain quantum effects can be observed~\cite{LipferSV15}.
	
	In the case of unbalanced homodyne detection, the signal state is also mixed with a coherent local oscillator beam at a beam splitter but only one \mbox{qPNRD} is recording one of the output modes.
	With such a setup it is possible to directly sample a click version of a s parametrized quasiprobability phase-space distribution~\cite{LuisSV15}
	\begin{equation}
		\label{eq:Pclick}
		P_{N} \left ( \alpha ; s \right ) = \frac{2}{\pi (1-s)} \sum_{k=0}^N \left [ \frac{\eta_{\det} (1-s)-2}{\eta_{\det} (1-s)} \right ]^k
		c_k  \left ( \alpha ; \eta_{\det} \right ). 
	\end{equation}
	Here $\alpha$ is the coherent amplitude of the local oscillator beam and the $c_k(\alpha ; \eta_{\det})$ are the recorded elements of the click counting statistics given the detector efficiency $\eta_{\det}$.
	Negativities in $P_{N} \left ( \alpha ; s \right )$ directly indicate quantum properties of the signal light field.
	It is obvious {from Eq.~\eqref{eq:Pclick}} that the knowledge about the detectors quantum efficiency $\eta_{\det}$ is crucial for this approach.
	Hence, a characterization of the detector response function is indispensable.

\section{Conclusion}\label{sec:conclusion}
	In this paper, we have presented a calibration method based on the click moments for \mbox{qPNRDs}. 
	We have applied this method to click statistics measured with a superconducting nanowire system and a time-bin multiplexing setup.
	By doing so, we have demonstrated that our method can compete with existing calibration methods, yet being very resource efficient, both from the experimental and the theoretical point of view.

	Furthermore, we found a strong polarization dependency of the of the detector response function.
	In particular, we showed how to account for polarization mixing effects, due to non polarization maintaining optical components in the setup, by interpreting the retrieved response function as an average over all response functions of the individual bins.
	This effect may be circumvented by using only polarization maintaining components.
	Finally, we proposed applications of well characterized \mbox{qPNRDs} for sensing turbulent atmospheric channels and performing phase sensitive measurements.

\section*{Acknowledgements}
	The project leading to this application has received funding from the European Union’s Horizon 2020 research and innovation programme under Grant No. 665148.
	R.K., J.S., C.S., and W.V. acknowledge support via the European Union under Grant No. 665148 (QCUMbER).
	M.B. and W.V. are grateful for financial support by Deutsche Forschungsgemeinschaft through Grant No. VO 501/22-1. 

%
%


\begin{thebibliography}{99}
	\bibitem{Nielsen} 
		M. A. Nielsen and I. L. Chuang, 
		\textit{Quantum Computation and Quantum Information} 
		(Cambridge University Press, Cambridge, 2000).
	\bibitem{Giovannetti2011} 
		V. Giovannetti, S. Lloyd, and L. Maccone, 
		``Advances in quantum metrology'',
		Nat. Photon. \textbf{5}, 222 (2011).
	\bibitem{Gisin2007} 
		N. Gisin and R. Thew, 
		``Quantum communication'',
		Nat. Photon. \textbf{1}, 165 (2007).
	\bibitem{NTH12}
		C. M. Natarajan, M. G. Tanner, and R. H. Hadfield,
		``Superconducting nanowire single-photon detectors: physics and applications'',
		Supercond. Sci. Technol. \textbf{25}, 063001 (2012).
	\bibitem{WDSBY04}
		E. Waks, E. Diamanti, B. C. Sanders, S. D. Bartlett, and Y. Yamamoto,
		``Direct Observation of Nonclassical Photon Statistics in Parametric Down-Conversion'',
		Phys. Rev. Lett. \textbf{92}, 113602 (2004).
	\bibitem{ABA10}
		A. Allevi, M. Bondani, and A. Andreoni,
		``Photon-number correlations by photon-number resolving detectors'',
		Opt. Lett. \textbf{35}, 1707 (2010).
	\bibitem{PDFEPW10}
		G. Puentes, A. Datta, A. Feito, J. Eisert, M. B. Plenio, and I. A. Walmsley,
		``Entanglement quantification from incomplete measurements: applications using photon-number-resolving weak homodyne detectors'',
		 New J. Phys. \textbf{12}, 033042 (2010).
	\bibitem{DYSTS11}
		J. F. Dynes, Z. L. Yuan, A. W. Sharpe, O. Thomas, and A. J. Shields,
		``Probing higher order correlations of the photon field with photon number resolving avalanche photodiodes'',
		Opt. Express \textbf{19}, 13268 (2011).
	\bibitem{MMDL12}
		P.-A. Moreau, J. Mougin-Sisini, F. Devaux, and E. Lantz,
		``Realization of the purely spatial Einstein-Podolsky-Rosen paradox in full-field images of spontaneous parametric down-conversion'',
		Phys. Rev. A \textbf{86}, 010101(R) (2012).
	\bibitem{DBIMHCE13}
		L. Dovrat, M. Bakstein, D. Istrati, E. Megidish, A. Halevy, L. Cohen, and H. S. Eisenberg,
		``Direct observation of the degree of correlations using photon-number-resolving detectors'',
		Phys. Rev. A \textbf{87}, 053813 (2013).
	\bibitem{DBJVDBW14}
		G. Donati, T. J. Bartley, X.-M. Jin, M.-D. Vidrighin, A. Datta, M. Barbieri, and I. A. Walmsley,
		``Observing optical coherence across Fock layers with weak-field homodyne detectors'',
		Nature Commun. \textbf{5}, 5584 (2014).
	\bibitem{FJPF03}
		M. J. Fitch, B. C. Jacobs, T. B. Pittman, and J. D. Franson,
		``Photon-number resolution using time-multiplexed single-photon detectors'',
		Phys. Rev. A \textbf{68}, 043814 (2003).
	\bibitem{ASSBW03}
		D. Achilles, C. Silberhorn, C. \'Sliwa, K. Banaszek, and I. A. Walmsley,
		``Fiber-assisted detection with photon number resolution'',
		Opt. Lett. \textbf{28}, 2387 (2003).
	\bibitem{RHHPH03}
		J. \v{R}eh\'a\v{c}ek, Z. Hradil, O. Haderka, J. Pe\v{r}ina Jr., and M. Hamar,
		``Multiple-photon resolving fiber-loop detector'',
		Phys. Rev. A \textbf{67}, 061801(R) (2003).
	\bibitem{SVA12}
		J. Sperling, W. Vogel, and G. S. Agarwal,
		``True photocounting statistics of multiple on-off detectors'',
		Phys. Rev. A \textbf{85}, 023820 (2012).
	\bibitem{SVA12a}
		J. Sperling, W. Vogel, and G. S. Agarwal,
		``Sub-Binomial Light'',
		Phys. Rev. Lett. \textbf{109}, 093601 (2012).
	\bibitem{SVA13}
		J. Sperling, W. Vogel, and G. S. Agarwal,
		``Correlation measurements with on-off detectors'',
		Phys. Rev. A \textbf{88}, 043821 (2013).
	\bibitem{BDJDBW13}
		T. J. Bartley, G. Donati, X.-M. Jin, A. Datta, M. Barbieri, and I. A. Walmsley,
		``Direct Observation of Sub-Binomial Light'',
		Phys. Rev. Lett. \textbf{110}, 173602 (2013).
	\bibitem{SBVHBAS15}
		J. Sperling, M. Bohmann, W. Vogel, G. Harder, B. Brecht, V. Ansari, and C. Silberhorn,
		``Uncovering Quantum Correlations with Time-Multiplexed Click Detection'',
		Phys. Rev. Lett. \textbf{115}, 023601 (2015). 
	\bibitem{HSPGHNVS15}
		R. Heilmann, J. Sperling, A. Perez-Leija, M. Gr\"afe, M. Heinrich, S. Nolte, W. Vogel, and A. Szameit,
		``Harnessing click detectors for the genuine characterization of light states'',
		Sci. Rep. \textbf{6}, 19489 (2016).
	\bibitem{SBDBJDVW16}
		J. Sperling, T. J. Bartley, G. Donati, M. Barbieri, X.-M. Jin, A. Datta, W. Vogel, and I. A. Walmsley,
		``Quantum correlations from the conditional statistics of incomplete data'',
		Phys. Rev. Lett. \textbf{117}, 083601 (2016).
	\bibitem{KK64}
		P. L. Kelley and W. H. Kleiner, 
		``Theory of Electromagnetic Field Measurement and Photoelectron Counting'',
		Phys. Rev. \textbf{136}, A316 (1964).
	\bibitem{VW06}
		W. Vogel and D.-G. Welsch, 
		Quantum Optics 
		(Wiley-VCH, Weinheim, Germany, 2006).
	\bibitem{tomography1} 
		A. Luis and Sanchez-Soto,
		``Complete characterization of arbitrary quantum measurement processes'',
		Phys. Rev. Lett. \textbf{83}, 3573 (1999).
	\bibitem{tomography2} 
		J. Fiur\'{a}\v{s}ek,
		``Maximum-likelihood estimation of quantum measurement'',
		Phys. Rev. A \textbf{64}, 024102 (2001).
	\bibitem{tomography3} 
		G. M. D'Ariano, L. Maccone, and P. Lo Presti, 
		``Quantum calibration of measurement instrumentation'',
		Phys. Rev. Lett. \textbf{93}, 250407 (2004).
	\bibitem{tomography4} 
		J. S. Lundeen, A. Feito, H. Coldenstrodt-Ronge, K. L. Pregnell, Ch. Silberhorn, T. C. Ralph, J. Eisert, M. B. Plenio, and I. A. Walmsley
		``Tomography of quantum detectors'',
		Nat. Phys. \textbf{5}, 27 (2009).
	\bibitem{tomography5} 
		A. Feito, J. S. Lundeen, H. Coldenstrodt-Ronge, J. Eisert, M. B. Plenio, and I. A. Walmsley,
		``Measuring measurement: theory and practice'',
		New J. Phys. \textbf{11}, 093038 (2009).
	\bibitem{tomography6} 
		P. C. Humphreys, B. J. Metcalf, T. Gerrits, T. Hiemstra, A. E. Lita, Jo. Nunn, S. W. Nam, A. Datta, W. S. Kolthammer, and I. A. Walmsley,
		``Tomography of photon-number resolving continuous-output detectors'',
		New J. Phys. \textbf{17}, 103044 (2015).
	\bibitem{tomographyqPNRD1} 
		H. B. Coldenstrodt-Ronge, J. S. Lundeen, K. L. Pregnell, A. Feito, B. J. Smith, W. Mauerer, Ch. Silberhorn, J. Eisert, M. B. Plenio, and I. A. Walmsley,
		``A proposed testbed for detector tomography'',
		J. Mod. Opt \textbf{56}, 432 (2009).
	\bibitem{tomographyqPNRD2} 
		V. D'Auria, N. Lee, T. Amri, C. Fabre, and J. Laurat, 
		``Quantum Decoherence of Single-Photon Counters'',
		Phys. Rev. Lett. \textbf{107}, 050504 (2011).
	\bibitem{tomographyNW1} 
		M. K. Akhlaghi, A. H. Majedi, and J. S. Lundeen,
		``Nonlinearity in single photon detection: modeling and quantum tomography'',
		Opt. Express \textbf{19}, 21305 (2011).
	\bibitem{tomographyNW2} 
		J. J. Renema, G. Frucci, Z. Zhou, F. Mattioli, A. Gaggero, R. Leoni, M. J. A. de Dood, A. Fiore, and M. P. van Exter,
		``Modified detector tomography technique applied to a superconducting multiphoton nanodetector'',
		Opt. Express \textbf{20}, 2806 (2012).
	\bibitem{tomographyNW3} 
		C. M. Natarajan, L. Zhang, H. Coldenstrodt-Ronge, G. Donati, S. N. Dorenbos, V. Zwiller, I. A. Walmsley, and R. H. Hadfield,
		``Quantum detector tomography of a time-multiplexed superconducting nanowire single-photon detector at telecom wavelengths'',
		Opt. Express \textbf{21}, 893 (2013).
	\bibitem{twinbeam1} 
		S. V. Polyakov and A. L. Migdall,
		``High accuracy verification of a correlated-photon-based method for determining photon-counting detection efficiency'',
		Opt. Express \textbf{15}, 1390 (2007).
	\bibitem{twinbeam2} 
		A. P. Worsley, H. B. Coldenstrodt-Ronge, J. S. Lundeen, P. J. Mosley, B. J. Smith, G. Puentes, N. Thomas-Peter, and I. A. Walmsley,
		``Absolute efficiency estimation of photon-number-resolving detectors using twin beams'',
		Opt. Express \textbf{17}, 4397 (2009).
	\bibitem{twinbeam3} 
		 J. Pe\v{r}ina, Jr. O. Haderka, V. Mich\'alek, and M. Hamar,
		 ``Absolute detector calibration using twin beams'',
		 Opt. Lett. \textbf{37}, 2475 (2012).
	\bibitem{twinbeam4} 
		 J. Pe\v{r}ina, Jr., O. Haderka,  A. Allevi, and M. Bondani,
		 ``Absolute calibration of photon-number-resolving detectors with an analog output using twin beams'',
		 Appl. Phys. Lett. \textbf{104},  041113 (2014). 
	\bibitem{Klyshko}
		D. N. Klyshko, 
		``Use of two-photon light for absolute calibration of photoelectric detectors'',
		Sov. J. Quantum Electron. \textbf{10}, 1112 (1980).
	\bibitem{SVA14}
		J. Sperling, W. Vogel, and G. S. Agarwal,
		Quantum state engineering by click counting,
		Phys. Rev. A {\bf 89}, 043829 (2014). 
	\bibitem{Usenko} 
		V. C. Usenko, B. Heim, C. Peuntinger, C. Wittmann, C. Marquardt, G. Leuchs, and R. Filip, 
		``Entanglement of Gaussian states and the applicability to quantum key distribution over fading channels'',
		New J. Phys. \textbf{14}, 093048 (2012).
	\bibitem{Elser} 
		D. Elser, T. Bartley, B. Heim, C. Wittmann, D. Sych, and G. Leuchs,
		``Feasibility of free space quantum key distribution with coherent polarization states'',
		New J. Phys. \textbf{11}, 045014 (2009).
	\bibitem{Semenov2012} 
		A. A. Semenov, F. T\"{o}ppel, D. Yu. Vasylyev, H. V. Gomonay, and W. Vogel,
		``Homodyne detection for atmosphere channels'',
		Phys. Rev. A \textbf{85}, 013826 (2012).
	\bibitem{SVA15} 
		J. Sperling, W. Vogel, and G. S. Agarwal,
		``Balanced homodyne detection with on-off detector systems: Observable nonclassicality criteria'',
		Europhys. Lett. {\bf 109}, 34001 (2015).
	\bibitem{LuisSV15} 
		A. Luis, J. Sperling, and W. Vogel,
		``Nonclassicality Phase-Space Functions: More Insight with Fewer Detectors'',
		Phys. Rev. Lett. {\bf 114}, 103602 (2015). 
	\bibitem{LipferSV15} 
		T. Lipfert, J. Sperling, and W. Vogel,
		``Homodyne detection with on-off detector systems'',
		Phys. Rev. A {\bf 92}, 053835 (2015).
	\bibitem{Sudarshan}
		E. C. G. Sudarshan,
		Equivalence of Semiclassical and Quantum Mechanical Descriptions of Statistical Light Beams, 
		Phys. Rev. Lett. {\bf 10}, 277 (1963).
	\bibitem{Glauber} 
		R. J. Glauber, 
		Coherent and incoherent states of the radiation field, 
		Phys. Rev. {\bf 131}, 2766 (1963).
	\bibitem{SSTT13}
		H. Shibata, K. Shimizu, H. Takesue, and Y. Tokura, 
		``Superconducting Nanowire Single-Photon Detector with Ultralow Dark Count Rate Using Cold Optical Filters'',
		Appl. Phys. Express \textbf{6}, 072801 (2013).
	\bibitem{regression} 
		M. Krystek and M. Anton,
		``A weighted total least-squares algorithm for fitting a straight line'',
		Meas. Sci. Technol. \textbf{18}, 3438 (2007).
	\bibitem{polarisation} 
		S. N. Dorenbos, E. M. Reiger, N. Akopian, U. Perinetti, V. Zwiller, T. Zijlstra, and T. M. Klapwijk,
		``Superconducting single photon detectors with minimized polarization dependence'',
		Appl. Phys. Lett. \textbf{93}, 161102 (2008).
	\bibitem{Semenov2009} 
		A. A. Semenov and W. Vogel, 
		``Quantum light in the turbulent atmosphere'',
		Phys. Rev.  A \textbf{80}, 021802(R) (2009).
	\bibitem{beamwandering} 
		D. Yu. Vasylyev, A. A. Semenov, and W. Vogel, 
		``Toward Global Quantum Communication: Beam Wandering Preserves Nonclassicality'',
		Phys. Rev. Lett. \textbf{108}, 220501 (2012).
	\bibitem{VSV2016} 
		D. Yu. Vasylyev, A. A. Semenov, and W. Vogel, 
		``Atmospheric Quantum Channels with Weak and Strong Turbulence'',
		Phys. Rev. Lett. \textbf{117}, 090501 (2016).
	\bibitem{BSSV16} 
		M. Bohmann, A. A. Semenov, J. Sperling, and W. Vogel, 
		``Gaussian entanglement in the turbulent atmosphere'',
		Phys. Rev. A {\bf 94}, 010302(R) (2016). 
	\bibitem{BSSV16b}
		M. Bohmann, J. Sperling, A. A. Semenov, and W. Vogel, 
		``Higher-order nonclassical effects in fluctuating-loss channels'',
		arXiv:1610.09053 [quant-ph]. 

\end{thebibliography}
\end{document}